\documentclass[pdflatex,sn-mathphys-num]{sn-jnl}

\usepackage{graphicx}%
\usepackage{multirow}%
\usepackage{amsmath,amssymb,amsfonts}%
\usepackage{amsthm}%
\usepackage{mathrsfs}%
\usepackage[title]{appendix}%
\usepackage{xcolor}%
\usepackage{textcomp}%
\usepackage{manyfoot}%
\usepackage{booktabs}%
\usepackage{algorithm}%
\usepackage{algorithmicx}%
\usepackage{algpseudocode}%
\usepackage{listings}%

\usepackage{graphicx}
\usepackage{float}
\usepackage{xcolor}
\usepackage{physics}
\usepackage{dcolumn}
\usepackage{bm}
\usepackage{hyperref}

\usepackage{orcidlink}
\usepackage{caption}
\usepackage{subcaption}

\begin{document}

\title{Quantum Hamiltonian Embedding of Images for Data Reuploading Classifiers}

\author*[1]{
\fnm{Peiyong}
\sur{Wang}
}
\email{peiyongw@student.unimelb.edu.au}

\author[3,1]{
\fnm{Casey R.}
\sur{Myers}
}
\email{casey.myers@unsw.edu.au}

\author[2]{
\fnm{Lloyd C. L.}
\sur{Hollenberg}
}
\email{lloydch@unimelb.edu.au}

\author*[1]{
\fnm{Udaya}
\sur{Parampalli}
}
\email{udaya@unimelb.edu.au}

\affil[1]{
\orgdiv{School of Computing and Information Systems, Faculty of Engineering and Information Technology},
\orgname{The University of Melbourne},
\orgaddress{
 \country{Australia}}
}

\affil[2]{
\orgdiv{School of Physics},
\orgname{The University of Melbourne},
\orgaddress{
\country{Australia}}
}

\affil[3]{
\orgname{Silicon Quantum Computing Pty Ltd, UNSW Sydney},
\orgaddress{
\country{Australia}}
}

\date{\today}

\abstract{

When applying quantum computing to machine learning tasks, one of the first considerations is the design of the quantum machine learning model itself. Conventionally, the design of quantum machine learning algorithms relies on the ``quantisation" of classical learning algorithms, such as using quantum linear algebra to implement important subroutines of classical algorithms, if not the entire algorithm, seeking to achieve quantum advantage through possible run-time accelerations brought by quantum computing. However, recent research has started questioning whether quantum advantage via speedup is the right goal for quantum machine learning \cite{Schuld2022-ll}. Research also has been undertaken to exploit properties that are unique to quantum systems, such as quantum contextuality, to better design quantum machine learning models \cite{Bowles2023-rs}. In this paper, we take an alternative approach by incorporating the heuristics and empirical evidences from the design of classical deep learning algorithms to the design of quantum neural networks. We first construct a model based on the data reuploading circuit \cite{DataReUploading-Perez-Salinas2020-gi} with the quantum Hamiltonian data embedding unitary \cite{QuantumDataEncodingFromBook-Schuld2021-ak}\footnote{Code available on \url{https://github.com/peiyong-addwater/HamEmbedding}.}. Through numerical experiments on images datasets, including the famous MNIST and FashionMNIST datasets, we demonstrate that our model outperforms the quantum convolutional neural network (QCNN)\cite{QCNN-Cong_2019} by a large margin (up to over 40\% on MNIST test set). Based on the model design process and numerical results, we then laid out six principles for designing quantum machine learning models, especially quantum neural networks.

}

\maketitle


\section{\label{sec:sec1-intro}Introduction}

As a key application area of quantum computing, quantum machine learning \cite{Biamonte2017-pg} has received considerable attention as an area that may achieve a potential quantum advantage compared to classical machine learning/deep learning algorithms through runtime acceleration. The quest for achieving such acceleration has become a standard motivation while developing quantum machine learning algorithms, evidenced by the use of efficient quantum subroutines that could accelerate linear algebra calculations, such as the quantum principal component analysis algorithm (qPCA), which involves calculations of the eigenvalues and eigenvectors of a covariance matrix by quantum phase estimation \cite{Lloyd2014-ga}.

Unlike principal component analysis and kernel methods, which are often referred to as statistical learning algorithms, neural networks, with their ability to discover hidden patterns in large-scale unstructured datasets such as image and natural language, have gained momnentum since the invention of AlexNet \cite{Krizhevsky2017-ch}, and have become the foundation of modern artificial intelligence applications such as ChatGPT-4 \cite{bubeck2023sparks}. However, since time complexity is rarely the first priority when designing novel deep neural network architectures, which often rely on intuition and even inspirations from biological neural networks, it becomes less obvious that quantum computing should find any advantage or utility in deep learning and AI. Although recent research has attempted to integrate properties that are unique to quantum systems, such as contextuality, into the design of quantum machine learning models for specific types of tasks that could lead to quantum advantage \cite{Bowles2023-rs}, few studies have taken the intuition behind successful deep learning models into account and how to integrate them into quantum machine learning models. In this paper, we aim to bridge this gap by bringing such intuition to the design of quantum machine learning models, especially quantum neural networks, via numerical experiments for the design of a quantum machine learning model for benchmarking image processing tasks.

Our main contributions in this paper are as follow:
\begin{itemize}
    \item Construction of a quantum classifier based on the quantum Hamiltonian embedding approach and the data reuploading circuit;
    \item Results from numerical experiments show that the proposed model could outperform the baseline quantum convolutional neural network model \cite{QCNN-Cong_2019};
    \item Based on the model design process and the numerical experiments, we lay out a set of principles for future quantum machine learning (QML) model design.
\end{itemize}
The results of our paper further emphasise the importance of heuristics during the design of quantum machine learning models, especially heuristics and empirical knowledge that found in the extensive classical deep learning literature.

This paper is organised as follows: in the rest of this section, we briefly introduce the relevant research in applying quantum machine learning to image processing, as well as common quantum data embedding approaches. In Section~\ref{sec:sec2-methods}, we propose a quantum classification model based on the data reuploading circuit \cite{DataReUploading-Perez-Salinas2020-gi} and quantum Hamiltonian embedding method \cite{QuantumDataEncodingFromBook-Schuld2021-ak}. In Section~\ref{sec:experiments}, we demonstrate the effectiveness of this model by evaluating the classification performance on different datasets, including the famous MNIST \cite{lecun2010mnist} and FashionMNIST \cite{xiao2017fashionmnist} datasets. In Section \ref{sec:discussion}, we discuss the results obtained through numerical experiments which inform our proposed six basic principles for the design of quantum neural networks.

\subsection{\label{sec:sec1.1-qml-image} QML for Image Processing}

As an illustration of quantum neural network (QNN) design, we consider one of the most important tasks in modern artificial intelligence -- image processing, including image classification, segmentation, and generation. Since the success of AlexNet \cite{Krizhevsky2017-ch} at the ImageNet Large-Scale Visual Recognition Challenge 2012 (ILSVRC 2012), deep neural networks, especially deep convolutional neural networks (CNN), have dominated image-related tasks. Recently, the vision transformer (ViT) \cite{vit-dosovitskiy2021image} and its variants are trending in image-related tasks due to its structural compatibility with large language models and the potential to build a single unified multi-modal model. An important step in the vision transformer is to cut the image into patches, following the same inductive bias as convolutional neural networks. From the history of image processing with deep neural networks, we can see that there is a central principle in the network architectures designed throughout the years, which is the locality of information and translation invariance. This is reflected in both the convolution kernels in CNNs and image patches in ViTs.

In the quantum context, a number of approaches have been developed as a direct analogue to the classical CNN, namely the quantum convolutional neural network (QCNN) \cite{QCNN-Cong_2019}, which borrows the idea of localised operators, shared parameters and down-sampling from its classical counterpart. It has been benchmarked for binary classification with classical image data \cite{Hur2022-ad, Gong2024-iq} and its effectiveness demonstrated through experiments. Variational circuits other than QCNNs can also be applied to image processing, but often require classical methods to reduce the dimension of the input data \cite{QUANTUMSSL-Jaderberg2022-gh, Khatun2024-rp}. A popular choice is to use a pre-trained classical neural network, such as ResNet \cite{he2015deep}, to preprocess the original images and extract features \cite{zaman2024comparativeanalysishybridquantumclassical, khatun2024quantum}. This approach often involves a quantum-classical hybrid neural network, where the output layer of the classical neural network is replaced with a parameterised quantum circuit. However, the necessity of such an approach remains unclear, as the last layer of a classical neural network has a great similarity to logistic regression, which, by itself, is a simple machine learning model. When using a classical neural network for dimension reduction, the ``heavy lifting" of feature extraction is off-loaded to the classical neural network, and the extracted features are often classified by a simple machine learning model. There is also research that involves the implementation or mimicking of classical convolutional operations via quantum circuits, such as \cite{kerenidis2019quantum}, in which the quantum version of convolution is achieved through local unitary operators with parameter sharing, and pooling is achieved by tracing out a subset of qubits. Other research aims to replace only the convolution operation in a classical neural network, such as the quanvolutional neural network \cite{QuanvolutionalNN-Henderson2020-yw} and its variants \cite{Riaz2023-iy}. The data reuploading classifier \cite{DataReUploading-Perez-Salinas2020-gi} has also been adapted for image classification, such as \cite{SingleQubitDataReuploadingClassifier-Easom-Mccaldin2021-rh}, in which the pixel data in the image are encoded as rotation parameters together with trainable parameters that are shared among different patches of the same image. From this we can we can see that, when applying quantum machine learning to high-dimensional data, especially image data, it is a common practice to 
\begin{itemize}
    \item use a classical model to reduce the dimension of the original image data and extract task-related features, such as in \cite{khatun2024quantum} and \cite{zaman2024comparativeanalysishybridquantumclassical};
    \item use amplitude embedding to reduce the number of qubits required, such as in \cite{west2023drastic};
    \item use the data reuploading circuit or use a small circuit that only operates on localised patches of the image to reduce the number of qubits required when angle embedding is involved, such as in \cite{SingleQubitDataReuploadingClassifier-Easom-Mccaldin2021-rh} and \cite{Riaz2023-iy}.
\end{itemize}
For the remainder of this section, we will give a brief introduction to common quantum data embedding methods.


\subsection{\label{sec:sec1.2-QuantumDataEncoding}Quantum Data Embedding}

One of the most important steps when applying quantum machine learning to classical data is loading the data into the quantum computer. For example, it is difficult to find the classical counterpart of quantum data embedding for the loading of images, where classically images can easily be stored as rank-3 tensors and matrices in computer memory. Appropriate data embeddings are crucial to the success of both classical and quantum machine learning models. Similar data embedding processes, where the original data structure is not suitable to be directly processed by the machine learning model (mostly neural networks), could be found in research and applications that involve graph and natural language data. In each case, the data embedding method, as well as the machine learning model that follows data embedding, need to reflect the intrinsic properties of the data. Such intrinsic properties can sometimes be described simply as translation, rotation, and permutation symmetry \cite{Heredge2024-ds}. However, most of the time it lies more on a semantic level and is hard to describe via mathematical relations. 

There are three widely adopted data embedding methods for quantum machine learning \cite{QuantumDataEncodingFromBook-Schuld2021-ak}:

\begin{itemize}
    \item \textsc{Basis Embedding}. In basis embedding, a length$-n$ binary string is directly embedded as one of the basis states of an $n-$qubit quantum system by applying Pauli-X operators to the quantum bits that are supposed to encode the classical bit ``1". For example, to encode the binary bit string `0101' as a quantum state, one only needs to apply Pauli-X gates to the initial state $\ket{0000}$ on the second and fourth qubits:
    \begin{equation}
        0101_{2} \mapsto \ket{0101} = X_2 X_4 \ket{0000};
    \end{equation}
    \item \textsc{Angle Embedding}. In angle embedding, classical floating-point data is embedded as rotation angles of parameterized quantum gates, such as the Pauli rotation gates $R_X$, $R_Y$ and $R_Z$. For $\mathbf{x} = (x_1, x_2)^T$, one could use angle embedding to encode $\mathbf{x}$ as:
    \begin{equation}
        \mathbf{x} \mapsto R_X (x_1) R_Z(x_2)\ket{0};
    \end{equation}
    \item \textsc{Amplitude Embedding}. In amplitude embedding, the normalised padded data vector $\mathbf{x}=(x_0, x_1, \cdots, x_{2^N-1})^T$ is embedded as a quantum state of an $N-$ qubit system with real amplitudes: 
    \begin{equation}
        \mathbf{x}\mapsto \ket{\mathbf{x}} = \sum_{i=0}^{2^N-1} x_i\ket{i}.
    \end{equation}
\end{itemize}

There are also special embedding methods designed for image data, such as the flexible representation of quantum images (FRQI) \cite{FlexibleRepQuantumImage-Le2011-wi, QuantumImageRepSurvey-Yan2016-rf}, which embeds the data in a quantum state that takes spatial information into account. Since these methods still require specific amplitudes for the embedded quantum states, they could be viewed as an extension of the amplitude embedding method. To make amplitude embedding feasible with current quantum hardware, several approximate heuristic-based state preparation methods have been proposed, such as the GASP algorithm \cite{Creevey2023-fc}, in which the authors applied genetic algorithms to discover relatively low-depth quantum circuits for approximate state preparation, as well as the variational circuit-based approach to approximately prepare the FRQI states shown in \cite{shen2024classification}.

In this paper, we combine the quantum Hamiltonian embedding method, described in Section~\ref{sec:sec2.2-hamembedding}, and the data reuploading approach, described in Section~\ref{sec:sec2.1-dataReuploading}, for image classification tasks. Compared to angle embedding, embedding the image as a Hamiltonian puts the pixels in the image on a more equal footing, meaning that they go through the same type of mathematical operation. Also, the matrix representation of a quantum Hamiltonian for a qubit system is naturally ``two-dimensional", in the sense that it has the same shape as an (grey-scale) image, making it more suitable for modelling images.

\section{\label{sec:sec2-methods}Data Reuploading Classifier with Quantum Hamiltonian Embedding}

In this section we will discuss the quantum neural network classifier model based on the discussions from the previous section. We opt for Hamiltonian image embedding (Section~\ref{sec:sec2.2-hamembedding}) for data encoding and the data reuploading classifier \cite{DataReUploading-Perez-Salinas2020-gi} (Section~\ref{sec:sec2.1-dataReuploading}), for both their simplicity in terms of implementation with linear algebra libraries such as JAX \cite{jax2018github}, and the intuitive connection with classical neural networks for the data reuploading circuit. While looking for quantum advantage in terms of training and inference speedups is a legitimate aim, it is not the primary goal here. Instead, we seek to integrate heuristics from deep learning into quantum machine learning model design.

In the context of quantum computing, a quantum Hamiltonian can be written as a square matrix. This representation provides the opportunity to encode image data in a two-dimensional way, rather than flattening the image and/or using the pixel values as rotation angles of parameterised gates, which could introduce unwanted bias on the decision boundary. Also, the possibility of encoding an entire image as a quantum Hamiltonian with (polynomial) less qubits required than amplitude embedding and angle embedding gives us the chance to reduce classical preprocessing to minimum. We will see in Eqn.~\ref{eqn:emb-expand} that encoding an image as a quantum Hamiltonian provides a richer nonlinearity compared to that provided by quantum measurements, which could further enhance the expressitivity of our model.

\subsection{\label{sec:sec2.2-hamembedding}Hamiltonian Image Embedding}

As mentioned in the previous section, one of the most important components of a quantum machine learning model is how to embed classical data into the quantum computer. Since we are working with image data, there is preference for data embedding methods that preserve two-dimensional structures of images and transform image pixels with the same non-linearity function (activation functions) to avoid unwanted bias on the decision boundaries.

In this paper, we adopt the Hamiltonian embedding method \cite{QuantumDataEncodingFromBook-Schuld2021-ak, Yang2023-pc} for image data encoding. First, we ``Hermitianise" our square, greyscale, real-valued image matrix $M$ by:
\begin{equation}
    H_M = \frac{M+M^T}{2}.
\end{equation}
This is the only classical preprocessing required for our model, in addition to padding the image with zeros for the MNIST and FashionMNIST datasets. Here, the embedding unitary for the input image data is simply the matrix exponentiation of $H_M$:
\begin{equation}
    W(t;M) = e^{\frac{-{i}H_M t}{2}},
\end{equation}
where $t$ is a trainable parameter instead of the physical time. If we expand $W(t;M)$ in a Taylor series, we have:
\begin{equation}\label{eqn:emb-expand}
    W(t;M) = 1-\frac{{i}H_M t}{2^1}-\frac{H_M^2 t^2}{2!\times 2^2}+\frac{{i} H_M^3 t^3}{3!\times 2^3 }+ \cdots
\end{equation}
We can see that by simply time-evolving the (Hermitianised) image, a (matrix) polynomial function is applied on the whole image level, bring ``cheapter" nonlinearity compared to angle embedding with single-parameter rotation gates. Later in Section~\ref{sec:experiments}, we demonstrate that with the quantum Hamiltonian embedding approach, our model could outperform QCNN for various datasets. In our model, the Hamiltonian embedding of image $M$, parameterised by a single parameter $t$, will act as the data encoding unitary for our quantum machine learning model, which will be discussed in the following subsections.

\subsection{\label{sec:sec2.1-dataReuploading}Data Re-Uploading}

The data reuploading variational quantum circuit, first proposed in \cite{DataReUploading-Perez-Salinas2020-gi}, is derived from the guiding principles from classical neural networks that the data is reused multiple times in classical deep neural networks. Variational circuits representing quantum versions of neural networks can be written as (before measurement):
\begin{equation}\label{eqn:common-vqc}
    \ket{\psi(\mathbf{x};\boldsymbol{\theta})} = V(\boldsymbol{\theta})U_{\boldsymbol{\phi}}(\mathbf{x})\ket{0}^{\otimes n},
\end{equation}
where $V(\boldsymbol{\theta})$ are the variational layers parameterised by $\boldsymbol{\theta}$, and could be absorbed in the measurement observables $O$, becoming $\boldsymbol{O}(\boldsymbol{\theta})=V^\dagger(\boldsymbol{\theta})OV(\boldsymbol{\theta})$:
\begin{equation}
\begin{split}
        \bra{\psi(\mathbf{x};\boldsymbol{\theta})}O\ket{\psi(\mathbf{x};\boldsymbol{\theta})} &= \bra{0}^{\otimes n}U^\dagger_{\boldsymbol{\phi}}(\mathbf{x}) V^\dagger(\boldsymbol{\theta})OV(\boldsymbol{\theta})U_{\boldsymbol{\phi}}(\mathbf{x})\ket{0}^{\otimes n}\\
        &=\bra{0}^{\otimes n}U^\dagger_{\boldsymbol{\phi}}(\mathbf{x}) \boldsymbol{O}(\boldsymbol{\theta}) U_{\boldsymbol{\phi}}(\mathbf{x})\ket{0}^{\otimes n},
\end{split}
\end{equation}
$\mathbf{x}$ is the input data, and $U_{\boldsymbol{\phi}}(\mathbf{x})$ is the data encoding unitary, and could be parameterised with some other set of parameters $\boldsymbol{\phi}$. In this form, the input data only appears once in the model, while in classical neural networks, one input neuron can be accessed by more than one neuron in the hidden layer. Motivated by this difference, the data reuploading circuit can be written as follows:
\begin{equation}\label{eqn:reupload}
    \ket{\Psi(\boldsymbol{x}; \Vec{\boldsymbol{\omega}})}=\prod_{i=1}^L \left[V({\boldsymbol{\omega}}_i)U_{\boldsymbol{\phi}}(\boldsymbol{x})\right]\ket{0}^{\otimes n}.
\end{equation}
In this definition the data encoding unitary $U_{\boldsymbol{\phi}}(\mathbf{x})$, together with the parameterised layer $V$, are repeated $L$ times with the same data encoding unitary but with different parameters for $V$, $\Vec{\boldsymbol{\omega}}=\{{\boldsymbol{\omega}}_1, {\boldsymbol{\omega}}_2,\cdots,{\boldsymbol{\omega}}_L\}$. Also it has been proved that data reuploading circuits in principle exhibit a quantum advantage in terms of function approximation \cite{Yu2023-vf}.

\subsection{\label{sec:sec2.3-themodel}The Model}

\begin{figure}[ht!]
    \centering
    \includegraphics[width=0.7\textwidth]{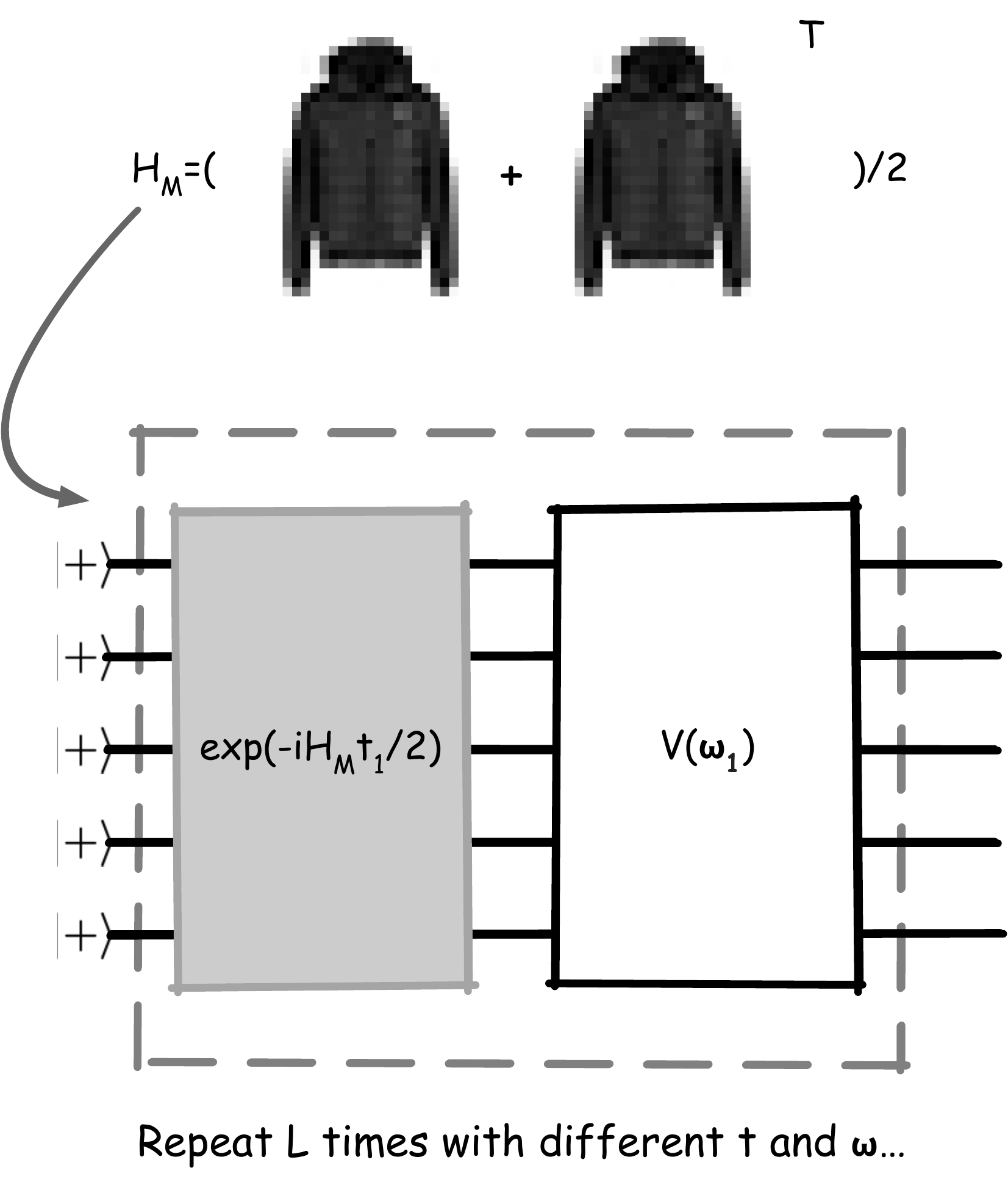}
    \caption{The quantum machine learning model described in Eqn.~\ref{eqn:actual-model}. The model shown in the figure has a five-qubit circuit for the images from the FashionMNIST dataset (and MNIST as well). The grey-scale images are padded from $28\times 28$ to $32\times 32$ with zeros. Then the quantum Hamiltonian $H_M$ is constructed with the padded image matrices. The data encoding unitary (grey box in the circuit diagram) and the parameterised circuit unitary (white box in the circuit diagram) are repeated $L$ times for an $L$-layered data re-uploading circuit.}
    \label{fig:ham_embed_model_full}
\end{figure}

Combining the data reuploading circuit and Hamiltonian embedding, we have the following quantum machine learning model (prior to measurement, also shown in Fig.~\ref{fig:ham_embed_model_full}):
\begin{equation}\label{eqn:actual-model}
    \begin{split}
        \ket{\varphi(\boldsymbol{t}, \Vec{\boldsymbol{\omega}};M)}&= \prod_{i=1}^L\left[V(\boldsymbol{\omega}_i)W(t_i;M) \right]\ket{+}^{\otimes n}.
    \end{split}
\end{equation}
Here we set the circuit to begin with an equal superposition of all basis states ($\ket{+}^{\otimes n}$). 

There is flexibility in the structure of the parameterised layer $V$. For small datasets, we opt for a parameterised layer composed of SU(4) unitary gates in a brick wall layout with different parameters. Generally, a SU(N) gate, where $N=2^n$, $n$ being the number of qubits the gate acts on, can be written as:

\begin{equation}
    \mathrm{SU}(N)(\boldsymbol{\theta}) = \exp (\sum_{i=1}^m \mathrm{i} \theta_i G_i), 
\end{equation}
where $m=4^n-1$ and $G_i\in\{I, X, Y, Z\}^{\otimes n}\backslash \{I^{\otimes n}\}$. $\boldsymbol{\theta} = \{\theta_1, \cdots, \theta_{4^n-1}\}$.

In our model, the classical data pass through a nonlinear operation first (time evolution), then followed by a parameterised unitary layer. Although this is not common in deep learning practice, where nonlinear operations (activation functions) normally occur after linear and convolutional layers, it could be viewed as a form of pre-activation, which enabled training of a 1001-layer ResNet in \cite{He2016-ks}.

\begin{figure}[ht!]
\centering
    \begin{subfigure}[c]{0.8\textwidth}
        \includegraphics[width=\textwidth]{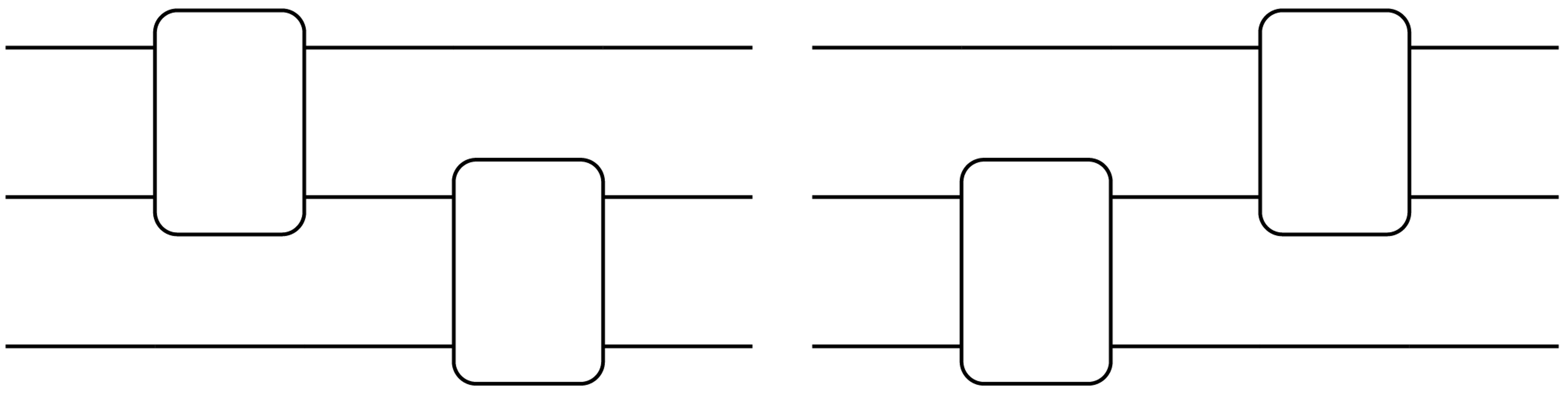}
        \caption{Two possible layouts for two SU(4) gates on a three-qubit circuit.}
        \label{fig:three-qubit-brickwall}
    \end{subfigure}

    \begin{subfigure}[c]{0.8\textwidth}
        \includegraphics[width=\textwidth]{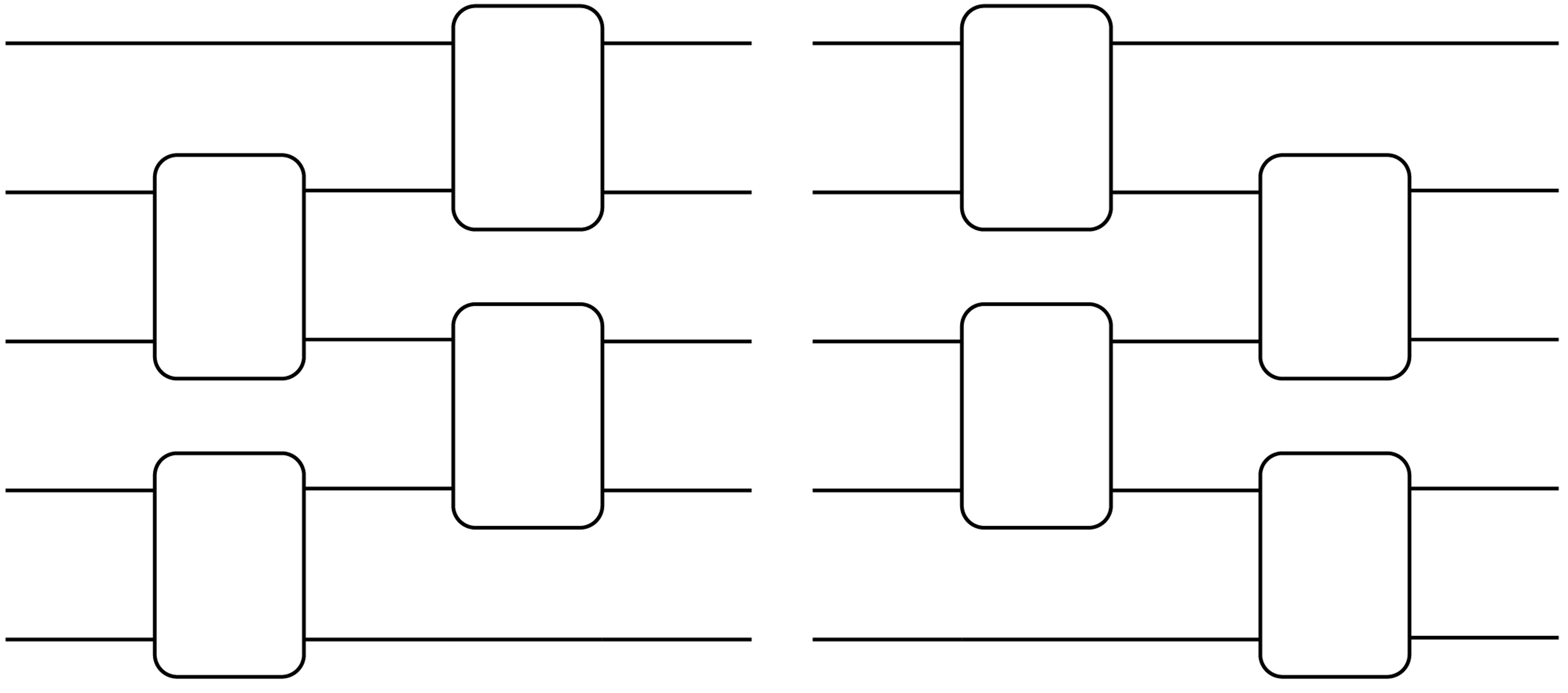}
        \caption{Two possible layouts for four SU(4) gates on a five-qubit circuit.}
        \label{fig:five-qubit-brickwall}
    \end{subfigure}
    \caption{Different layouts for the SU(4) gate in three- and five-qubit circuits.}
        \label{fig:su4layouts}
\end{figure}

For a three-qubit circuit, there could be two different configurations, as shown in Figure~\ref{fig:three-qubit-brickwall}. These two different configurations generally do not have a noticeable impact on the performance of the model. The same holds for the SU(4) gates in a five-qubit circuit, as shown in Figure~\ref{fig:five-qubit-brickwall}. For larger datasets, the SU(N) gate on all the qubits in the circuit will be adopted.

For the classification of $K-$ classes, the probability for each label $i\in \{0, 1, \cdots, K-1\}$ can be obtained by measuring the $\lceil \log_2 K \rceil- $ qubit projection operator $P_i = \ket{i}\bra{i}$:
\begin{equation}
    p(M;i) =\bra{\varphi(\boldsymbol{t}, \Vec{\boldsymbol{\omega}};M)} (P_i\otimes \boldsymbol{I}_{n-\lceil \log_2 K \rceil})\ket{\varphi(\boldsymbol{t}, \Vec{\boldsymbol{\omega}};M)},
\end{equation}where $\boldsymbol{I}_{n-\lceil \log_2 K \rceil}$ is the $(n-\lceil \log_2 K \rceil)-$qubit identity operator, and $n$ is the total number of qubits in the circuit. The loss function for training is the cross-entropy cost function:
\begin{equation}
    \text{Cross-Entropy Loss }(M)=-\sum_{i=0}^{K-1}y_i\log_2 p(M;i),
\end{equation}
where $y_i$ is the true probability of class $i$ for the input $M$, which, in the case of one-hot encoding, is 1 for the true class and 0 for all others. For a more detailed explanation of the cross-entropy function, readers could refer to deep-learning-related textbooks, such as Chapter 5.7 in \cite{prince2023understanding}.

To avoid taking the log of 0, we use $\text{Softmax}(p(M;i))$ to replace $p(M;i)$:
\begin{equation}
    \text{Softmax}(p(M;i)) = \frac{e^{p(M;i)}}{\sum_{k=0}^{K-1}e^{p(M;k)}}.
\end{equation}
The purpose of the Softmax function is to convert a real-valued vector to another real-valued vector, but with values in $(0,1)$ and sum to one\cite{prince2023understanding}.

\section{\label{sec:experiments} Simulation Experiments and Results}

\subsection{\label{sec:baseline-model}Baseline Model and Datasets} 
Our baseline model for comparison is the quantum convolutional neural network proposed in \cite{QCNN-Cong_2019}. The structure and implementation of the baseline model follow \cite{KorbinianKottmann2022}. For the baseline model, since amplitude embedding is used, all input images were flattened into vectors and padded. The size and number of parameters of the model depend on the size of the input and the number of classes.

\begin{figure}[ht!]
\centering
    \begin{subfigure}[b]{0.98\columnwidth}
        \includegraphics[width=\textwidth]{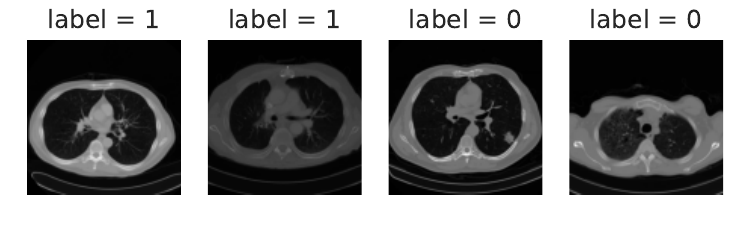}
        \caption{}
        \label{fig:ctimgoriginal}
    \end{subfigure}
    \vfill
    \begin{subfigure}[b]{0.98\columnwidth}
        \includegraphics[width=\textwidth]{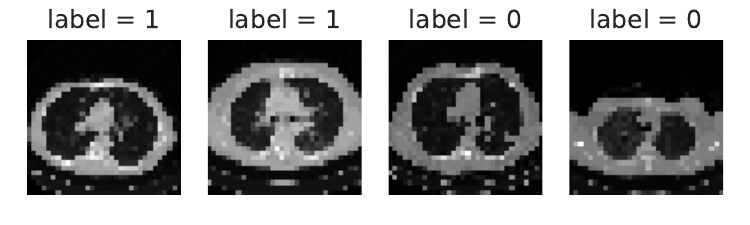}
        \caption{}
        \label{fig:ctimgresized}
    \end{subfigure}
    \caption{Sample images from the Kaggle CT Medical Image dataset. ~\ref{fig:ctimgoriginal}: Original image samples from the CT medical image dataset. ~\ref{fig:ctimgresized}: Same images as Fig.~\ref{fig:ctimgoriginal}, but resized to 32 by 32. The resize is achieved via OpenCV-Python's resize function. The pixels are also normalised before being fed into the quantum neural networks using OpenCV-Python's normalize function.}
        \label{fig:ctmedsample}
\end{figure}

We trained and tested our model on four different datasets:

\paragraph{The Kaggle CT Medical Images dataset} 
This is a small subset of images from \cite{ctimagefull}, obtained from the Kaggle website \cite{Scott_Mader2017-ch}. This dataset contains 100 CT medical images that have binary labels ``True" or ``False" for ``Contrast". The original dimension of the images is 512 by 512. To reduce the simulation cost, the images were resized to 32 by 32 using the Python package of OpenCV \cite{itseez2015opencv}. The resized images were randomly divided into train (80 images) and test (20 images) datasets. Sample images of the dataset are shown in Fig.~\ref{fig:ctmedsample}.
    
\begin{figure}[ht!]
    \centering
    \includegraphics[width=\textwidth]{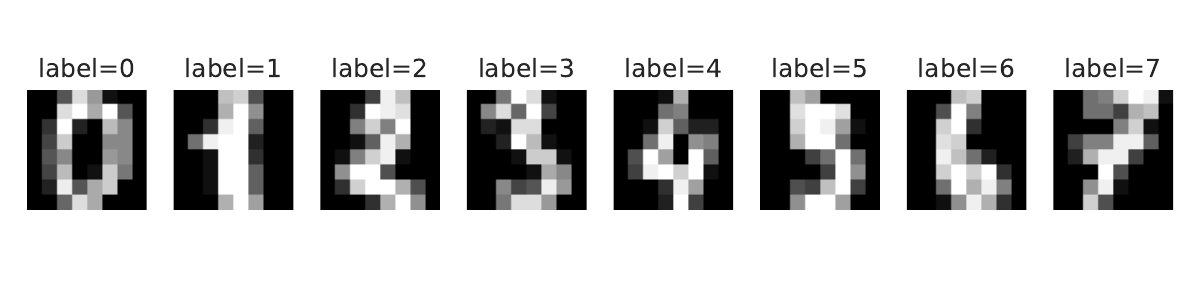}
    \caption{Sample images from the Sklearn digits dataset. The size of the images is eight by eight.}
    \label{fig:sklearn-digits}
\end{figure}

\paragraph{Subset of the Sklearn digits dataset} 
This data \cite{misc_optical_recognition_of_handwritten_digits_80} is obtained through the machine learning package ``Scikit-learn" \cite{scikit-learn}. The dimensions of the images are 8 by 8. Only images with labels 0 to 7 (eight classes in total) were sampled when splitting train (1200 images) and test (100 images) datasets. Sample images are shown in Fig.~\ref{fig:sklearn-digits}.

\begin{figure}[ht!]
    \centering
    \includegraphics[width=\textwidth]{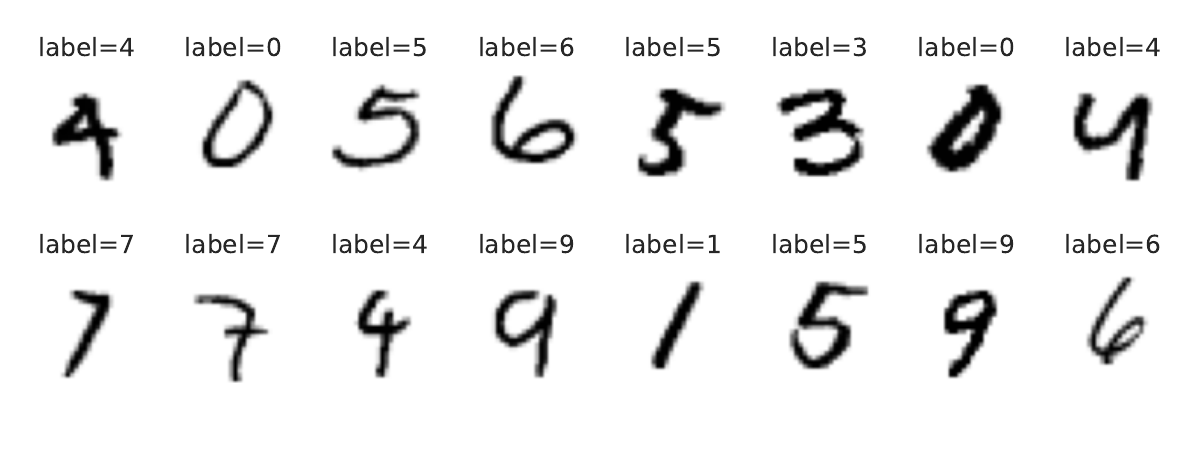}
    \caption{Sample images from the MNIST dataset. The size of the original images is 28 by 28. Images were padded with zeros to 32 by 32 before constructing the Hermitian operators of the images.}
    \label{fig:mnist}
\end{figure}

\paragraph{Subset of the MNIST dataset}  
The data \cite{lecun2010mnist} is obtained through the corresponding data loading module in Torchvision \cite{torchvision2016}. Only images with labels 0 to 7 (eight classes in total) were sampled when constructing the train datasets (48200 images) and the test datasets (8017 images). The original dimension of the images in the MNIST dataset is 28 by 28. The images were padded to 32 by 32 with zeros. Image samples are shown in Fig.~\ref{fig:mnist}.

\begin{figure}[ht!]
    \centering
    \includegraphics[width=\textwidth]{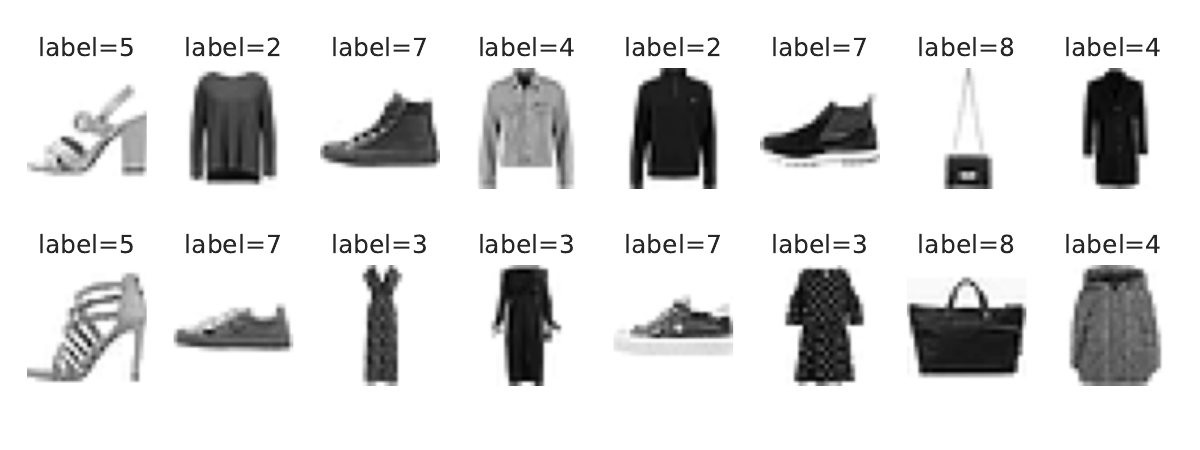}
    \caption{Sample images from the FashionMNIST dataset. The size of the original images is 28 by 28. Images were padded with zeros to 32 by 32 before constructing the Hermitian operators of the images.}
    \label{fig:fashionmnist}
\end{figure}

\paragraph{Subset of the FashionMNIST dataset}  
Data \cite{xiao2017fashionmnist} is obtained through the corresponding data loading module in Torchvision \cite{torchvision2016}. Only images with labels 0 to 7 (eight classes in total) were sampled when constructing the train datasets (48000 images) and the test datasets (8000 images). The original dimension of the images in the FashionMNIST dataset is 28 by 28. The images were padded to 32 by 32 with zeros. Image samples are shown in Fig.~\ref{fig:fashionmnist}.

\subsection{\label{sec:results}Results}

Although our model does not provide speedups for either training or inference when compared to classical techniques, it still outperforms the baseline QCNN model with different random initialisations and without extensive hyperparameter search. These results show how machine learning models based on straightforward heuristics could easily achieve good performance, even outperforming famous quantum machine learning models and reach a level close to classical machine learning models.

\begin{table}[ht]
\caption{\label{tab:all-res}%
The performance of the baseline and proposed models at the last iteration on all the datasets. For the Kaggle CT Medical Images dataset, the model performance data is averaged over 20 different parameter initialisations; For the Digits dataset, the performance is also averaged over 20 different parameter initialisations. For the MNIST and FashionMNIST dataset, the performances are both averaged over 5 different parameter initialisations. For the FashionMNIST dataset, we also have additional results, in which we doubled the depth of the data reuploading circuit. However, we can see that the performance increase is only marginal comapared to the increase in the number of parameters.
}

\begin{tabular}{lcccc}
\toprule
\textbf{Dataset}&
\textbf{Metrics}&
\textbf{QCNN}&
\textbf{HamEmb}&
\textbf{HamEmb (2x depth)}\\
\midrule
\multirow{4}{*}{Medical Images}&\textrm{Train Loss} & 0.6571 & \textbf{0.5457} & -\\
&\textrm{Test Loss} & 0.6843 & \textbf{0.6318} &- \\
&\textrm{Train Accuracy} & 73.25\% & \textbf{83.75\%} & -\\
&\textrm{Test Accuracy} & 58.25\% & \textbf{67.75\%}  & -\\
\midrule
\multirow{4}{*}{Digits}&\textrm{Train Loss} & 1.9461 & \textbf{1.5454} & -\\
&\textrm{Test Loss} & 1.9493 & \textbf{1.5609} &- \\
&\textrm{Train Accuracy} & 79.77\% & \textbf{95.22\%} & -\\
&\textrm{Test Accuracy} & 78.95\% & \textbf{94.40\%}  & -\\
\midrule
\multirow{4}{*}{MNIST}&\textrm{Train Loss} & 1.9889 & \textbf{1.5742} & -\\
&\textrm{Test Loss} & 1.9871 & \textbf{1.5740} &- \\
&\textrm{Train Accuracy} & 46.94\% & \textbf{89.24\%} & -\\
&\textrm{Test Accuracy} & 47.03\% & \textbf{89.72\%}  & -\\
\midrule
\multirow{4}{*}{FashionMNIST}&\textrm{Train Loss} & 1.9738 & 1.6072 &\textbf{1.5810}\\
&\textrm{Test Loss} & 1.9741 & 1.6114 & \textbf{1.5876}\\
&\textrm{Train Accuracy} & 43.46\% & 78.52\% & \textbf{80.46\%}\\
&\textrm{Test Accuracy} & 42.90\% & 77.77\%  & \textbf{79.61\%}\\
\botrule
\end{tabular}

\end{table}

\begin{figure}[ht!]
    \centering
    \includegraphics[width=0.6\linewidth]{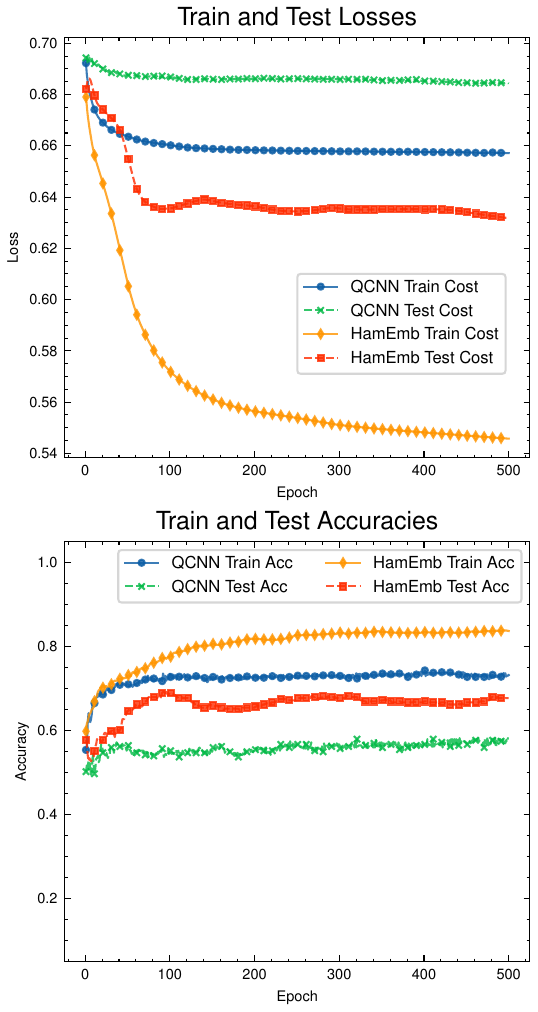}
    \caption{Plots for the averaged train and test metrics over 20 different parameter initialisations for the Kaggle CT Medical Images dataset. Although the proposed model (HamEmb) outperforms the baseline model, we could still see the gap between performances on the train and test dataset, potentially due to the small size of the dataset.}
    \label{fig:ct-plot}
\end{figure}

\begin{figure}[ht!]
    \centering
    \includegraphics[width=0.6\linewidth]{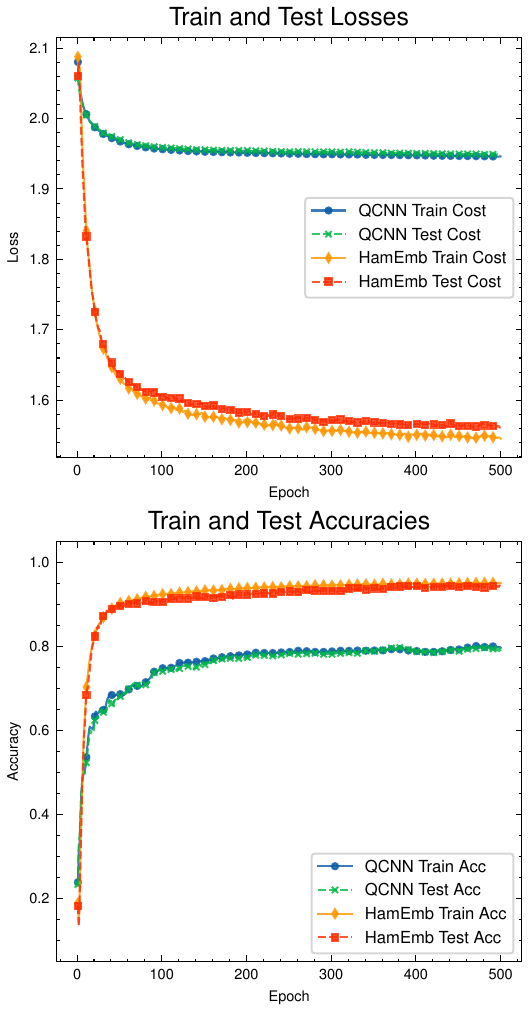}
    \caption{Plots for the averaged train and test metrics over 20 different parameter initialisations for the Sklearn digits dataset. We can see that the model proposed (HamEmb) drastically outperform the baseline model (QCNN with amplitude embedding).}
    \label{fig:digits-plot}
\end{figure}

\begin{figure}[ht!]
    \centering
    \includegraphics[width=0.6\linewidth]{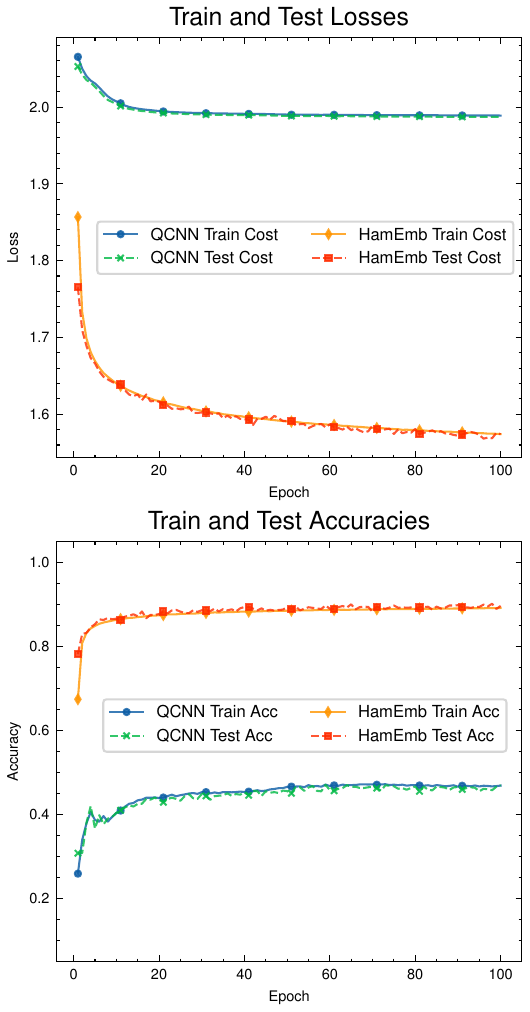}
    \caption{Plots for the averaged train and test metrics over 5 different parameter initialisations for the MNIST handwritten digits dataset. We can see that the model proposed (HamEmb) outperforms the baseline model (QCNN with amplitude embedding). However, we can observe that both the performances of the baseline and proposed model have dropped compared to the performance shown in Fig.~\ref{fig:digits-plot}, also for handwritten-digit-type dataset, which could potentially caused by the increased dimension of the image size in the dataset.}
    \label{fig:mnist-plot}
\end{figure}

\begin{figure}[ht!]
    \centering
    \includegraphics[width=0.6\linewidth]{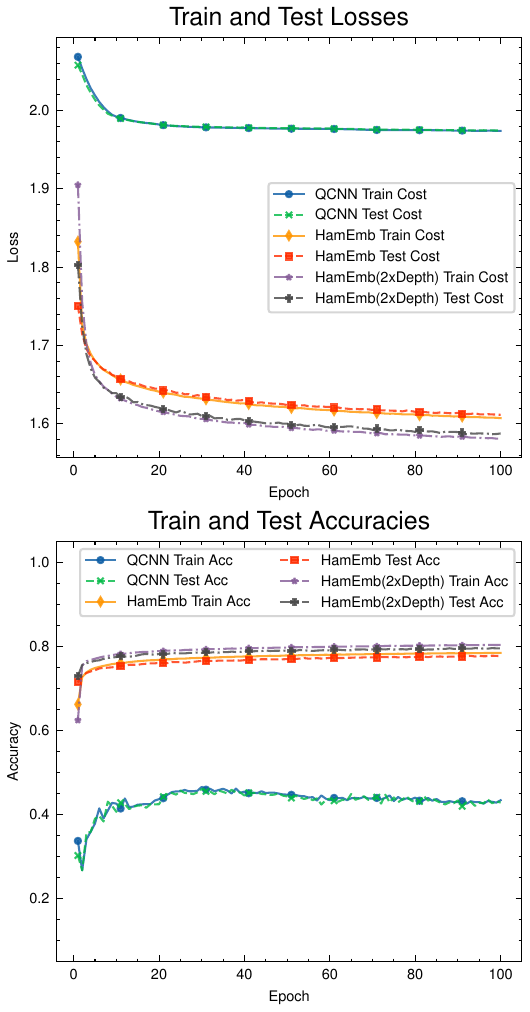}
    \caption{Plots for the averaged train and test metrics over five different parameter initialisations for the FashionMNIST dataset. We can see that the model proposed (HamEmb) outperforms the baseline model (QCNN with amplitude embedding). However, we can observe that both the performances of the baseline and proposed model have dropped compared to the performance shown in Fig.~\ref{fig:mnist-plot}, and increasing the depth two times does not increase the performance by a significant margin. This outcome could potentially be due to the decreased sparsity of the images in the FashionMNIST dataset compared to the MNIST dataset.}
    \label{fig:fashionmnist-plot}
\end{figure}

Both the baseline and the proposed model were trained with 20 different parameter initialisations for the Kaggle CT Medical Images dataset and subset of the Sklearn Digits dataset, each for 500 iterations with the Adam optimiser \cite{kingma2017adam}. In contrast, the MNIST and FashionMNIST subset datasets were trained for five different parameter initialisations, each for 100 iterations with the Adam optimiser \cite{kingma2017adam}.  For the CT Medical Images and Sklearn digits datasets the loss and accuracy are averaged over the 20 different parameter initialisations. For the MNIST and FashionMNIST datasets the loss and accuracy are averaged over the 5 different parameter initialisations. The averaged values of the evaluation metrics (loss and accuracy) of the final iteration for each of these four datasets can be found in Table 1, and the curve plot of the metrics through the training iterations could be found in Figs.~\ref{fig:ct-plot}, ~\ref{fig:digits-plot}, ~\ref{fig:mnist-plot} and ~\ref{fig:fashionmnist-plot}. A summary of results could be found in Table~\ref{tab:all-res}.


\subsection{\label{sec:analysis}Analysis}

The performance gaps for the three datasets Sklearn Digits, MNIST and FashionMNIST between the proposed models and the baseline models can clearly be seen in Figs.~\ref{fig:digits-plot}, ~\ref{fig:mnist-plot} and ~\ref{fig:fashionmnist-plot}. The performances of the proposed model on the test dataset are consistently better than those of the baseline model on the training dataset. However, the performance separation on the Kaggle CT Medical Images dataset is not as significant. A major difference between the Kaggle CT Medical Images dataset and the other three the datasets is that it requires downsampling (from $512\times 512$ to $32\times 32$) before data embedding. From Fig.~\ref{fig:ctimgoriginal} and Fig.~\ref{fig:ctimgresized}, we can see that the downsampling process obscured many fine-grained features in the image. This downsampling process, in addition to the limited number of samples in the dataset, could be one of the reasons that led to this  smaller performance difference between the proposed model and the baseline model.

Comparing the performance in the Sklearn Digits dataset and the MNIST dataset, we can see that although both the baseline model and the proposed model experienced a performance drop when the size of the input data increased, the performance of the baseline model dropped more and widened the gap between the baseline model and the proposed model. This phenomenon shows that the baseline model, which adopts amplitude embedding as the data encoding method, cannot efficiently capture the features in such a high dimensional image dataset. The performance drop of the proposed model, although smaller compared to that of the baseline model, still indicates that it could also have trouble dealing with high-dimensional image data, but is saved by the increased number of parameters.

When comparing the performance of our scheme on MNIST and FashionMNIST, the baseline model has a smaller performance drop compared to the proposed model, as shown in Fig.~\ref{fig:mnist-plot} and Fig~\ref{fig:fashionmnist-plot}, as well as Table~\ref{tab:all-res}, albeit still performing worse than the proposed model. FashionMNIST has richer and more complex spatial features compared to the MNIST dataset. The small change in performance of the baseline models indicates that it is likely to fail in effectively capturing the features in both datasets. The performance drop for our proposed model implies that we could deduce that the features in the FashionMNIST dataset are also more challenging compared to those in the MNIST dataset. By encoding the entire image as a quantum Hamiltonian, our model could potentially have characteristics similar to a convolution layer with a very large kernel. Large convolutional kernels cannot capture fine-grain details in the image as well as smaller sized convolutional kernels, and the lack of local features could be the reason the performance drops for both the baseline model and the proposed model.

\section{\label{sec:discussion}Discussion}

In the previous sections we designed a quantum neural network model based on the data reuploading circuit \cite{DataReUploading-Perez-Salinas2020-gi}, using the quantum Hamiltonian embedding \cite{QuantumDataEncodingFromBook-Schuld2021-ak} approach as the data encoding unitary, demonstrating that our model could achieve reasonably better performance than the well-known QCNN model without extensive architecture and hyperparameter search on multiple datasets, or dedicated pre-trained variational circuits to approximate quantum-embedded classical images \cite{Shen2024-me}.

It should also be noted that our numerical experiments use larger datasets compared to previous quantum machine learning research, since data scaling is also an important research question in different areas of deep learning, such as large language models \cite{kaplan2020scaling,hoffmann2022training}. It is common for machine learning models that have good performance on a small subset of common datasets, such as MNIST and FashionMNIST, to perform badly on a larger scale, both in terms of number of labels and number of datum in each label.

In this section we point out similarities between our model and classic neural network designs.  We followed heuristic techniques when choosing the data embedding methods and the structure of the QNN model. In this section we will dive into the details of these heuristics and propose six essential principles for quantum neural network design to inspire future research in this direction. 

\subsection{Resemblances to Classical Neural Network Design\label{sec:similar-to-classical-nn}}

\textbf{Possible connection to pre-activation in classical neural networks: } 
In \cite{He2016-ks} a modified version of the original ResNet \cite{he2015deep}, the activation function (ReLU), was placed before the convolutional layers (and after the batch normalisation layer). This modification \cite{He2016-ks} has been shown to increase the trainability of a 1000-layer ResNet and reduce overfitting. In the proposed model presented here, we can see from the expansion of the Hamiltonian embedding in Eqn.~\ref{eqn:emb-expand} that our training data first go through a non-linear transform, then a parameterised layer. This implicit transform of the input data could be the reason that during training, the proposed model did not suffer from any obvious barren plateau with different random initialisations. However, as this stage this is just a conjecture and still requires further investigation.

\quad

\noindent\textbf{Possible connection to the Gated Linear Unit \cite{dauphin2017language}: } Generally, the Gated Linear Unit (GLU) follows the form
\begin{equation}
    \operatorname{GLU}(x) = f(x)\cdot \sigma(g(x)),
\end{equation}
where both $f$ and $g$ are linear transformations (such as the linear layer in an MLP or the convolution layer in a CNN), $\sigma$ is usually a non-linear activation function, such as $\operatorname{ReLU}$ or the $\tanh$ function. Recall the mathematical form of our model from Eqn.~\ref{eqn:actual-model}:
\begin{equation}
    \begin{split}
        \ket{\varphi(\boldsymbol{t}, \Vec{\boldsymbol{\omega}};M)}&= \prod_{i=1}^L\left[V(\boldsymbol{\omega}_i)W(t_i;M) \right]\ket{+}^{\otimes n},
    \end{split}
\end{equation}
which can be rewritten as:
\begin{equation}
    v(\mathbf{x}; \boldsymbol{t},\boldsymbol{V}) = \prod_{i=1}^N [V_i\cdot \sigma_{t_i}(\mathbf{x})] v_0,
\end{equation}
where $V_i\cdot\sigma_{t_i}(\mathbf{x})$ can be viewed as a gated linear unit parameterised by weight matrix $V_i$ and parameter $t_i$. Although it is hard to say that there is a one-to-one correspondence between GLU and the proposed model, the implicit data transform in the quantum Hamiltonian embedding unitary could contribute to the superior performance of our model compared to other QCNN schemes.

We observed a performance degradation in the proposed model occurring when the complexity of the data increases (Sklearn digits $\rightarrow$ MNIST $\rightarrow$ FashionMNIST). An intuitive explanation is that the images in the FashionMNIST dataset contain more details than those in the MNIST dataset. Most pairs of MNIST digits have been shown to be well distinguished by using just a single pixel \cite{xiao2017fashionmnist}. Our result for the FashionMNIST brings us close to the performance achieved by classical machine learning algorithms on the benchmark provided in \cite{xiao2017fashionmnist}. 
Since we did not extensively search for better structures, there is ample room for improvement when it comes to model architecture. For example, our model encodes the entire image as a whole, whereas modern deep learning practice show that even without convolution, we should still divide the image into patches, allowing the model to focus more on local spatial features, such as the embedding layer in the vision transformer \cite{Dosovitskiy2020-kv}.

\subsection{\label{sec:principles}Principles of Quantum Machine Learning Model Design}

\begin{figure}[ht!]
    \centering
    \includegraphics[width=\textwidth]{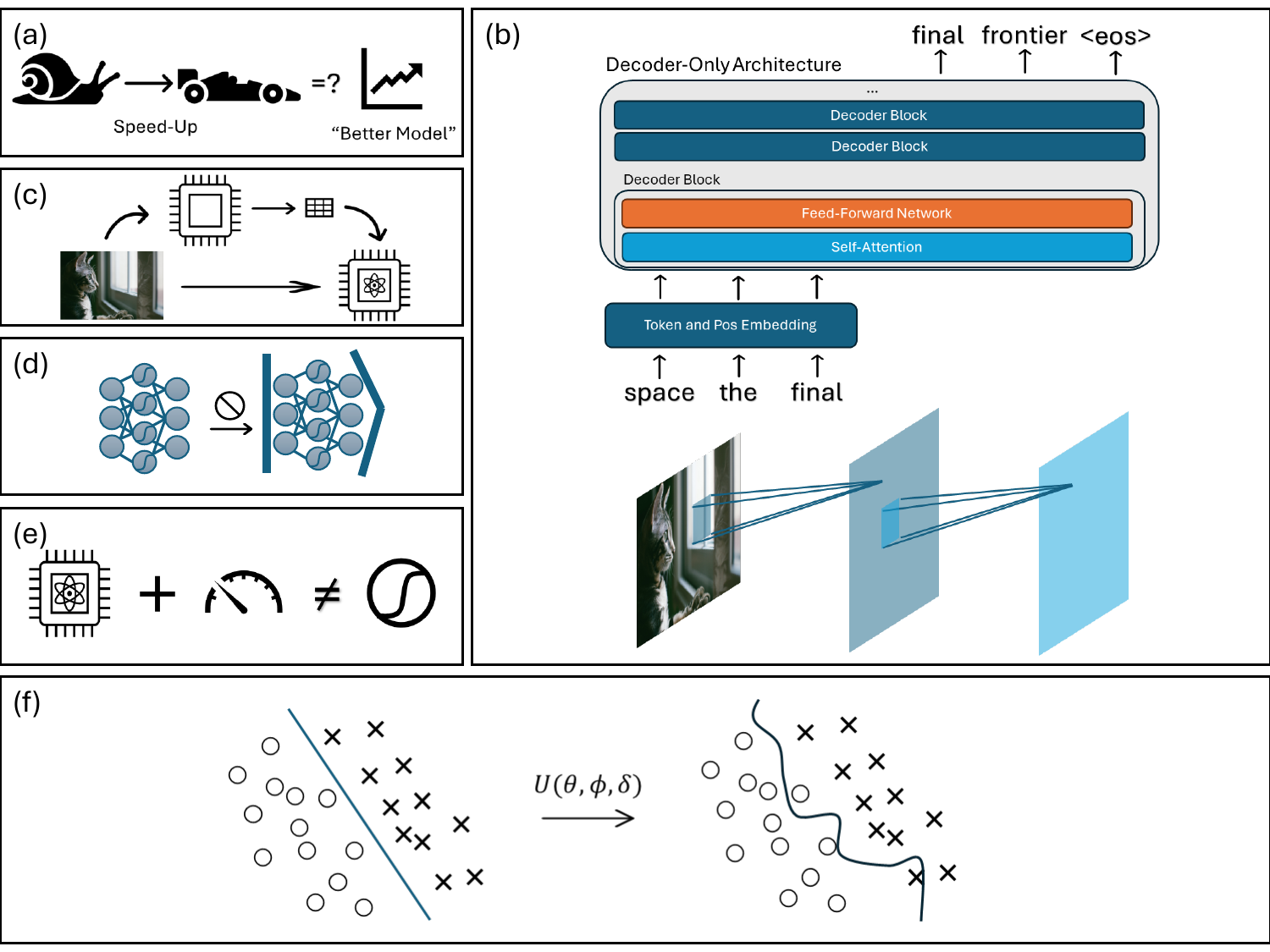}
    \caption{Our proposed six guiding principles required for designing quantum machine learning models. Box (a): Speedup is not the first thing to consider when designing new quantum machine learning models, since it does not necessarily lead to ``better models", i.e. we should instead focus in improvements in metrics of performance, such as accuracy. Box (b): The intrinsic structure of the data should be taken into account when designing the model architecture. Images have two major spatial directions, so the convolution kernel in a CNN (lower part of Box (b)) will scan in both directions, while text data only has a single temporal dimension, so the model needs to generate the words one by one (upper part of Box (b), which is a decoder-only generative transformer \cite{GPT1}). Box (c): Using a classical model, such a neural network backbone or PCA to reduce the dimension of the data (the upper path), obscures the real effectiveness of the quantum machine learning model, so we should minimise classical preprocessing as much as possible. Box (d): Avoid direct ``quantisation", i.e. avoid using quantum circuits to implement the exact mathematical operations of a classical machine learning model. Box (e): Measurements in the quantum circuit of the quantum machine learning model may not have the same kind of nonlinearity as the activation functions in classical neural networks, such as the ReLU (rectified linear unit, $\operatorname{ReLU}(x) = \operatorname{max}(0, x)$) function. Box (f): Some data embedding methods, such as angle embedding, may introduce unwanted bias toward certain kinds of decision boundaries, harming the performance of the machine learning model.}
    \label{fig:leading-fig}
\end{figure}

In this paper, we assume that, as with their classical counterparts, quantum machine learning with quantum neural network (QNN) models also need to process input data and to discover hidden patterns, which means that they may also benefit from incorporating similar intuition to that used for classical deep neural networks, while operating under the framework of quantum computing. In the previous sections, we briefly mentioned some of the reasons for our design choices of the QNN model. In this section, we lay out the following, but non-exhaustive, guiding principles for the design of QML models (shown schematically in Figure~\ref{fig:leading-fig}).

\begin{enumerate}

    \item \textsc{Less Initial Focus on Speedups}. During the conceptualisation of our model, we did not put speedup quantum advantage as the initial goal (Box (a) of Fig.~\ref{fig:leading-fig}). Instead, we searched for options that aligns with the structure of the data (which will be further discussed later), as well as using heuristics from classical neural networks (the data reuploading circuit). Since we have been demonstrated with numerical experiments that our model has better performance relative to QCNN results (Table~\ref{tab:all-res}), in the future, research could be focused on how to efficiently implement such quantum Hamiltonian embedding on current quantum hardware. Similar evolution has occurred in the research of deep learning models. For example, after the Transformer model proved its superior performance in language processing tasks \cite{Vaswani2017-es}, acceleration methods such as flash attention \cite{dao2022flashattention} were proposed to speed up the calculation of the attention layers. When developing quantum machine learning models, one could follow a similar principle: first find a framework that has acceptable performance on common datasets and then move onto the optimisation of the model.

    \item \textsc{Keep the Data in Mind}. One of the major reasons that the quantum Hamiltonian embedding approach, instead of many other popular data embedding methods, was selected during the model design process is that by embedding the images as quantum Hamiltonian in the form of matrices, we could preserve the two-dimensional structure of the images as much as possible. This particular choice is often referred as the inductive bias in machine learning literature. In classical machine learning, inductive bias is a set of assumptions a model makes to generalise better on unseen inputs, since there is no completely general learning algorithms according to the No Free Lunch theorem \cite{Goyal2022-bj}.
    This inductive bias is also integrated into the design of classical convolutional neural networks, where the convolution kernel usually moves along the height and width of the image to capture the spatial dependence of pixels on different locations (see Box (b) of Fig.~\ref{fig:leading-fig}). Even when the convolutional neural network is applied to sequence data \cite{kim2014convolutional}, the kernel usually moves along the time dimension to capture temporal correlations, which is different from the CNNs that operate on image data. In more recent research, transformer-based architectures \cite{Vaswani2017-es, GPT1} have often been used for language modelling. We can see from these two examples that even though the underlying operation (convolution) is the same, different kinds of data will require different kinds of convolution kernels.

    \item \textsc{Minimise Classical Preprocessing}. In our model, the only classical preprocessing steps involved are the normalisation of pixel values, which is also a common practice in classical deep learning, as well as the padding, transpose of and addition between the image matrices to construct the quantum Hamiltonian (see Box (c) in Fig.~\ref{fig:ham_embed_model_full}). It has been a common practice in some of the quantum machine learning research to adopt the backbone of a classical neural network, such as the ResNet-18 \cite{he2015deep}, to extract task-specific features from the high-dimensional image data (for an example, see \cite{zaman2024comparativeanalysishybridquantumclassical}). In essence, this kind of approach replaces the last fully-connected layer of the backbone classical neural network with a parameterised quantum circuit. Usually, for classification tasks, the activation function at the last layer (the fully-connected layer) is the softmax function, making the last layer a multinomial logistic regression. Figuratively speaking, we can say that the classical backbone neural network has done most of the ``heavy lifting" of the downstream task by extracting features from the image. Even a trivial (multinomial) logistic regression classifier could complete the task, which makes one question the necessity of introducing quantum models at the end of the classical model. In the design of our model, we avoided this potential issue by minimising the classical preprocessing as much as we could.

    \item \textsc{Avoid Direct ``Quantisation" of Classical Models}. Instead of directly ``quantising" the classical machine learning model to a quantum one which performs the same arithmetic operations on a quantum computer via quantum linear algebra, it would be preferable to develop quantum neural network models that could utilise operations that are intrinsic to the underlying quantum system, such as the time evolution of a quantum Hamiltonian, indicated by the numerical experiment results that the quantum Hamiltonian embedding, which is related to the time evolution of a quantum system, outperforms the QCNN, which ``quantise" the classical convolution and pooling operations via qubit-local unitary operators and measurement-outcome-controlled unitary operators. We could see similar trends happening in classical deep learning research: models that harvest most of the hardware prevail. One of the reasons that transformers \cite{Vaswani2017-es} have such advantage over recurrent-neural-network-based models such as the Long-Short Term Memory (LSTM) network and the Gated Recurrent Unit (GRU) network is that the structure of transformer models enables it to be easily parallelised during training and accelerated by GPUs. In other words, the structure of transformer models is more compatible with current GPU architectures.

    \item \textsc{Nonlinearity Is Not What You Think}. During the design of our model, we relied on the evolution of the image Hamiltonian to introduce ``nonlinearity" into the (quantum) neural network, which resembles the gated linear unit, as shown in Section~\ref{sec:similar-to-classical-nn}. However, in quantum computing, measurement operations are often considered as a ``nonlinear" operation since the the time evolution of the system is no longer defined by the Schr\"odinger equation, which is a reversible linear differential equation. While in deep learning research, nonlinearity refers to the ability of the model or a layer to nonlinearly (that is, not just translation, scaling, or rotation) transform the data manifold \cite{Olah_undated-bg}. Recent research shows that nonlinear functions (activation functions) play a more important role rather than merely providing nonlinear transformations on the data manifold. In \cite{dnnalwaysgrok}, the authors suggest that nonlinearity (ReLU) in deep neural networks divides the input space into non-overlapping linear regions. In \cite{teney2024neural}, the authors suggest that activation functions introduce a non-trivial bias to the neural network, making the neural network favour functions with certain levels of complexity. For example, ReLU-activated neural networks would favour low-complexity (low-frequency) functions which often align with the training target. Features learned by neural networks can also be regarded as directions in the activation space \cite{elhage2022superposition}. Since operations conditioned on mid-circuit measurement results could be converted to quantum controlled operations via deferred measurements, the transformations on the input states by the quantum convolutional neural network could be represented as a complete-positive trace preserving (CPTP) map, which is linear (but not reversible). Although quantum neurons with repeat-until-success circuits could produce nonlinear responses akin to a classical activation function \cite{cao2017quantumneuronelementarybuilding}, the input is first passed to an $R_Y$ gate, which is already nonlinear. The repeat-until-success model could be viewed as a filtering process to produce the tanh-like signal. In addition, the repeat-until-success process makes it difficult to scale to a large number of neurons, which is common for today's large models.

    \item \textsc{Be Careful of Unwanted Bias}. The other reason that we opted for quantum Hamiltonian embedding rather than popular choices, such as angle embedding, was to avoid unwanted bias. In recent research, the authors of \cite{Bowles2024-kb} showed that angle embedding for quantum machine learning models could introduce bias on decision boundaries that are formed based on periodic functions, which is also taken into account during the design of our model (Box (f) of Fig.~\ref{fig:leading-fig}), since the rotation angles passed as gate parameters will first go through trigonometric functions. This is essentially the same as using the sine or cosine functions as activation functions, which is very rare in deep learning practice. If parameterised gates, which take multiple rotation angles as input parameters, such as the U gate $\operatorname{U}(\theta, \phi, \delta)$, different elements (pixels) of the same datum (image) will go through different nonlinear operations. Unless we are sure that this is required due to the nature of the data or the task, we should avoid such different treatments to the same datum.

\end{enumerate}

In summary, our proposed principles for designing quantum machine learning models shed new light in how to combine quantum computing and machine learning for classical data. Some of these principles could be viewed as inductive biases, which enable the algorithm to prioritise certain hypotheses over others. Others are motivated by the underlying hardware that runs the model. However, it should be noted that these six principles are far from complete. The hardware-based principles could be extended to analogue neural networks on the quantum processors (or, in Hinton's words, ``mortal computation" \cite{Kleiner2024-xb} on a quantum processor). The principles for inductive biases could be in a difficult position in the future due to recent research in large language models, especially those on the scaling laws \cite{kaplan2020scaling, Hoffmann2022-wa}. According to Sutton \cite{Sutton_undated-gi}, one could say that the past 70 years of AI research could be summarised into the development of more and more general methods with weaker modelling assumptions or inductive biases, adding more data and compute power, or in other words, to scale up. There has been evidence that given enough data and compute budget, even MLPs (multi-layer perceptron) and the closely related MLP-Mixer models could perform in-context learning, which is the ability to solve a task from only input examples \cite{Tong2024-ld}. Also, in addition in \cite{Nguyen2024-og}, through numerical experiments with a pixel transformer that treats an image as a set of pixels and employs randomly initialised and learnable position embeddings without any information about 2D structure, the authors questioned the necessity of the inductive bias of locality which presents in many computer vision models, from LeNet \cite{LeCun1989-nw} to the vision transformer \cite{vit-dosovitskiy2021image}. It remains to be seen if such considerations impact the introduction of inductive biases from the quantum side to QML models \cite{Bowles2023-rs}. With limited computation, we still need inductive biases during the design of QML models, which should be comforting for researchers in this area. However, in the future, when (classical and/or quantum) compute is cheaper and more accessible for machine learning and AI research, enabling the ability to train even larger models with a larger amount of data, it would be necessary to remove such inductive biases from the design of the model \cite{Chung_2024}. 

It also should be noted that whether to continue investment of resources for the scaling law should be the future of AI research is still under heavy debate. There are still many issues and limitations existing in the current auto-regressive decoder-only transformer large language models \cite{Barbero2024-tx, Abbe2024-hv, Nezhurina2024-sk, Ofir_Press2022-pu, Verma2024-cl, Wu2024-xf, Kambhampati2024-gk, Zhou2023-hq}, and not all of them could be solved by scaling up the size of the data and computation resources invested in the training process, especially those regarding compositional reasoning \cite{Dziri2023-ws,Wang2024-tk}. In the context of current work in quantum computing, there is still a significant gap between the major concerns in today’s AI research and the research on quantum advantage and utility for AI. The research presented here seeks to find ways to bridge this gap.

\bigskip

\backmatter

\bmhead{Acknowledgements}

The authors acknowledge the support of IBM Quantum Hub at the University of Melbourne and the Seed Grant from School of Computing and Information Systems, the University of Melbourne.

\section*{Declarations}

\begin{itemize}
\item Funding: The authors received no outside funding during this project.
\item Data/Code availability: Code and data are available on \url{https://github.com/peiyong-addwater/HamEmbedding}.
\end{itemize}

\newpage
\appendix

\bibliography{ref}

\end{document}